\documentclass[toc]{PoS}
\usepackage{caption}
\usepackage{subcaption}
\usepackage{sidecap}
\sidecaptionvpos{figure}{c}
\usepackage{wrapfig}
\title{ Measurement of the Transverse Single-Spin Asymmetries for $\pi^{0}$ and Jet-like Events at Forward Rapidities at STAR in $p+p$ Collisions at $\sqrt{s}$ = 500 GeV}
\author{Mriganka Mouli Mondal (for the STAR Collaboration)\\
  Texas A\&M University, College Station, TX  77843, USA \\
  E-mail: \email{mriganka@rcf.rhic.bnl.gov}}
\abstract{Large transverse single-spin asymmetries ($A_N$) have been observed
  for forward inclusive hadron production in $p+p$ collisions at
  various experiments. In the collinear perturbative scattering
  picture, twist-3 multi-parton correlations can give rise to such an
  asymmetry.  A transversely polarized quark can also give rise to a
  spin-dependent distribution of its hadron fragments via the 
  higher twist effects  or the Collins fragmentation function. 
  The observed $A_N$ may involve contributions from both processes. 
  These can be disentangled by studying asymmetries for jets, direct
  photons and jet-fragments.\\
  The STAR Forward Meson Spectrometer (FMS), a Pb-glass electromagnetic
  calorimeter covering the pseudo-rapidity ($\eta$) range 2.6-4.2 and full
  azimuth, can detect photons, neutral pions and eta mesons.  We are
  measuring $A_N$ for $\pi^{0}$ and jet-like events reconstructed 
  from photons in the FMS in $p+p$ collisions at $\sqrt{s}$ = 500 GeV
  that were recorded during the 2011 RHIC run. We study $A_N$ as a function of 
  the number of observed photons in the FMS, thereby exploring asymmetries 
  for a range of event classes. We further study $A_N$ for forward jets 
  and its dependency on forward-midrapidity jet correlation. The current 
  status of the analysis will be discussed.
}
\FullConference{XXII. International Workshop on Deep-Inelastic Scattering and Related Subjects\\ Warsaw, 28 April - 2 May 2014} 
\ShortTitle{TSSA for $\pi^{0}$ and Jet-like Events at Forward Rapidities at STAR}
\begin{document}
\section{Introduction}
Transverse single-spin asymmetries (TSSAs), $A_{N} = \frac{\sigma^{\uparrow} - \sigma^{\downarrow}}{\sigma^{\uparrow} + \sigma^{\downarrow}}$, play an important role in understanding
the QCD structure of the nucleon. Surprisingly large TSSAs were
observed at forward rapidity for inclusive pions 
($p^{\uparrow}p{\rightarrow}{\pi}X$) in the 1970s and 1980s~(\cite{c1} and 
references therein). The large asymmetries seen in 
fixed-target experiments persist up to the center-of-mass energies 
available at RHIC ~\cite{c1a}.
There are two theoretical frameworks in QCD to explain the challenging large 
TSSAs. The first one relies on the transverse-momentum dependent (TMD) parton
distributions for a transversely polarized proton and spin depdent fragmentation 
known as Collins fragmentation function.  The Sivers function 
correlates a parton's transverse momentum to the proton's
spin, with large TSSAs as an outcome with the additional effect of Collins fragmentation~\cite{c2} . The other theoretical 
framework introduces the spin-dependent twist-three quark-gluon correlations
(Efremov-Teryaev-Qiu-Sterman (ETQS) mechanism)~\cite{c4} and twist-3
 fragmentation functions~\cite{c6a}. 
The TMD mechanism is relevant for two-scale ($p_T$ and $Q$) processes with 
$p_{T} \ll Q$, whereas the twist-3 approach is appropriate for single-scale processes
with $p_{T}$ providing the hard scale. 
There should be an
interesting overlapping region, $\Lambda_{QCD} \ll p_{T} \ll Q$, where both 
the theoretical pictures are valid. For the observables like inclusive $\pi^0$
and jets in {\it p+p} collisions at RHIC energies ($\sqrt{s}$ =
200 GeV or 500 GeV), the measurements involve one scale in high $p_T$, indicating
that the twist-3 approach is the more relevant description. An 
interesting feature in the latter description is still to observe the
expectation of 1/$p_T$ falling behavior of TSSAs at high $p_T$.
\section{Neutral Electromagnetic Jets and Analysis Procedure}
The Forward Meson Spectrometer (FMS) is a lead-glass electromagnetic calorimeter
in the STAR detector~\cite{c7} at RHIC consisting of an array of smaller 
cells ($3.8 \times 3.8$ cm$^2$) embedded in an array of larger cells ($5.8 \times 5.8$ cm$^2$). Sitting
in the very forward region, with 
pseudo-rapidity coverage $2.6 < \eta < 4.0$ and full
azimuthal acceptance, the FMS can detect neutral particles like $\pi^0$, $\eta$ mesons,
and $\gamma$. The large acceptance allows one to identify
electromagnetic jets (EM-Jets) in this important kinematic region where the 
asymmetries for inclusive hadron production are known to be large. The STAR central calorimetric system consists of
the Barrel Electromagnetic Calorimeter (BEMC) and Endcap Electromagnetic Calorimeter 
(EEMC) covering the pseudo-rapidity $-1.0 < \eta < 1.0$ and 
$1.1 < \eta < 2.0$, respectively.
By reconstructing EM-Jets in the forward and mid-rapidity calorimeters, 
we can study the dependence of the forward EM-Jet asymmetry on a central 
EM-Jet coincidence.
 The study of the $A_N$'s in the jet framework in addition to the 
earlier studies ~\cite{c8} is expected to give deeper insight into the origin
of the asymmetries related to the event topology.
\begin{figure}
  \centering
  \begin{subfigure}{.38\textwidth}
    \centering
    \includegraphics[width=1.0\linewidth]{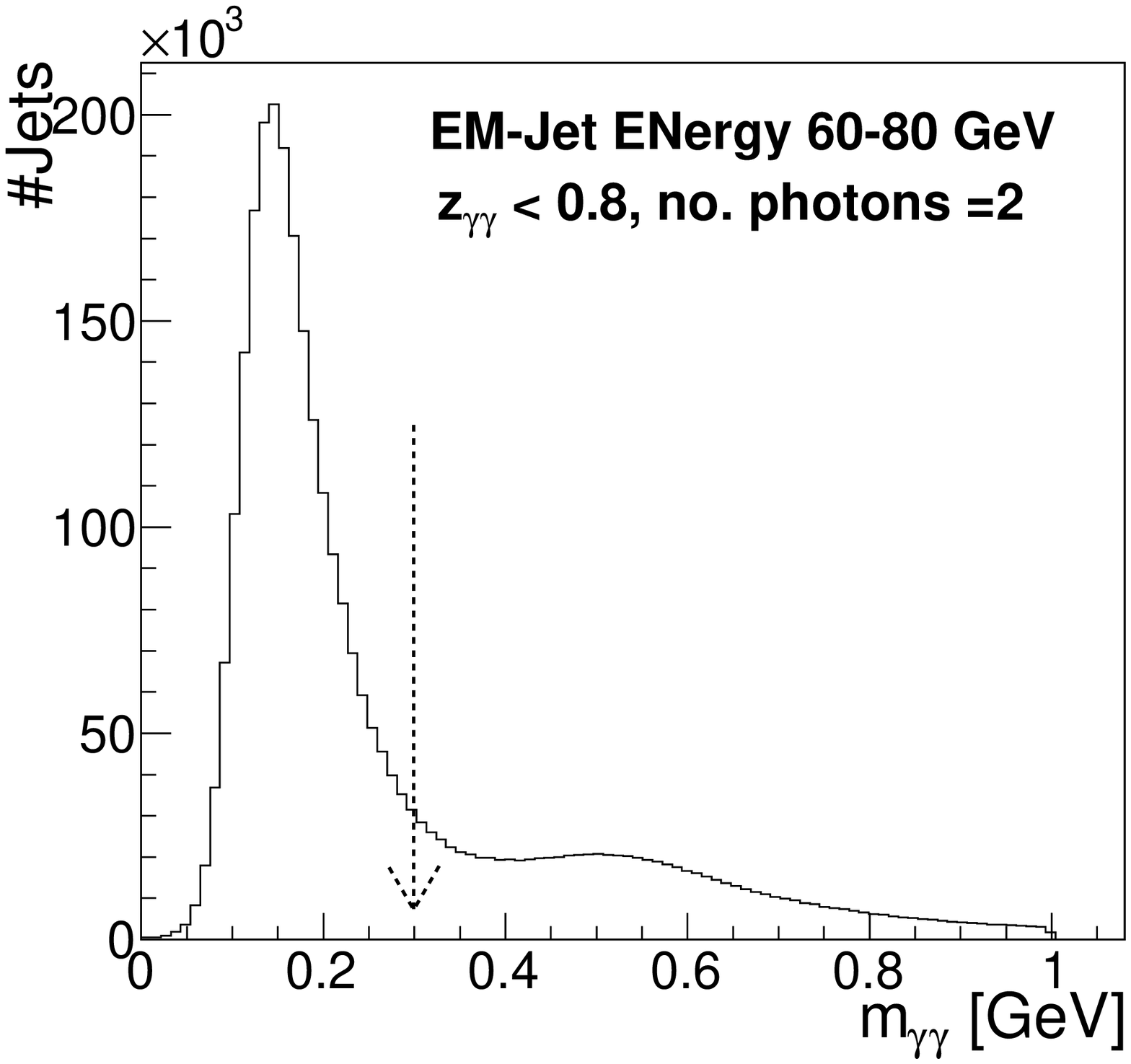}
    \caption{}
    \caption*{(a) 2$\gamma$ invariant mass distributions for EM-Jets containing two photons. $\pi^0$Jets are defined with $m_{\gamma\gamma}$ < 0.3 GeV/$c^2$.\\(b) Energy flow for forward jets within the jet cone for EM-Jets containing different numbers of photons.}
    \label{fig1}
  \end{subfigure}%
  \begin{subfigure}{0.42\textwidth}
    \centering
    \includegraphics[width=1.0\linewidth]{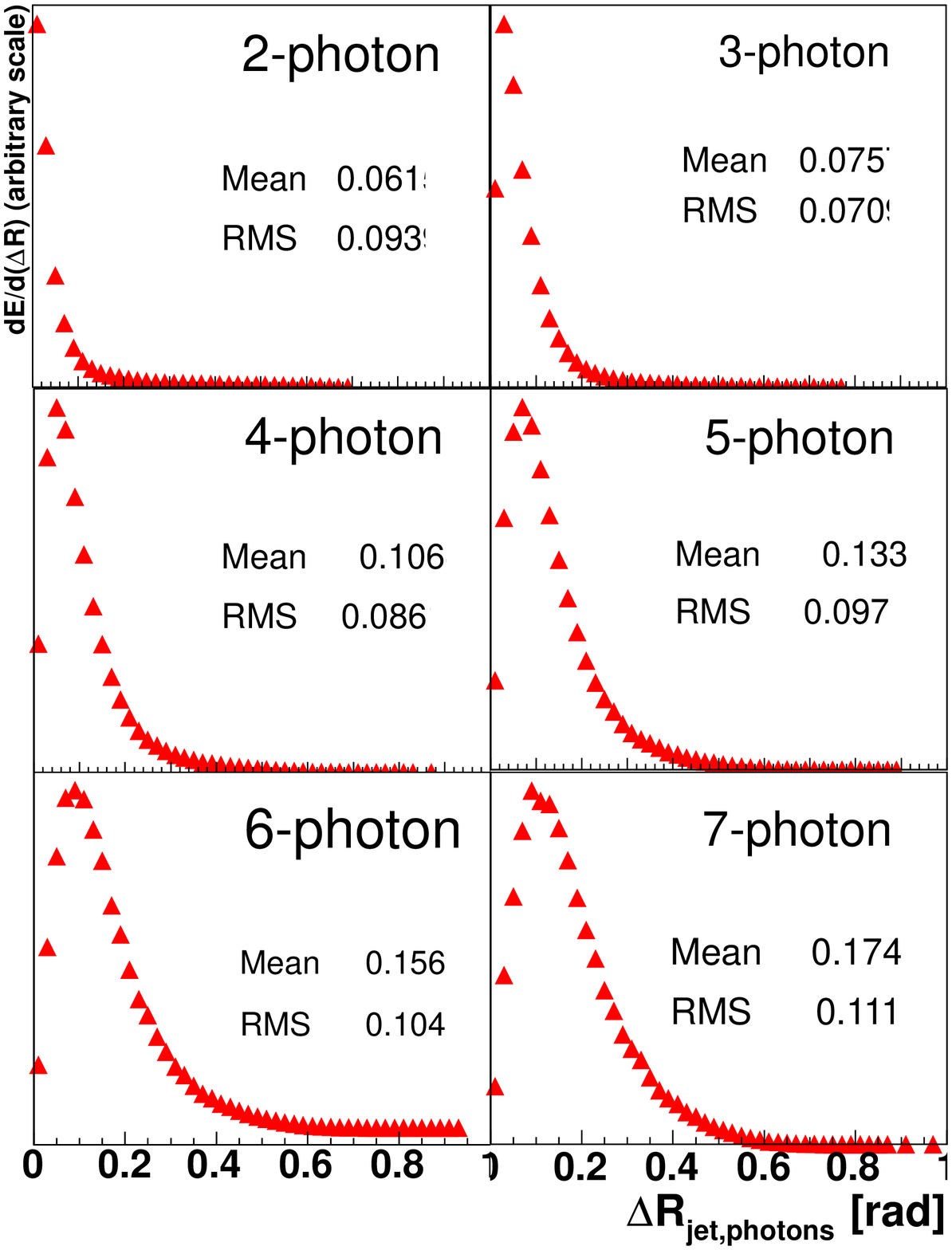}
    \caption{}
    \label{fig2}
  \end{subfigure}
  \label{F-one}
  \caption{}
\end{figure}
\begin{wrapfigure}{r}{0.46\textwidth}
  \centering
  \begin{subfigure}{.225\textwidth}
    \centering
    \includegraphics[width=1.0\linewidth]{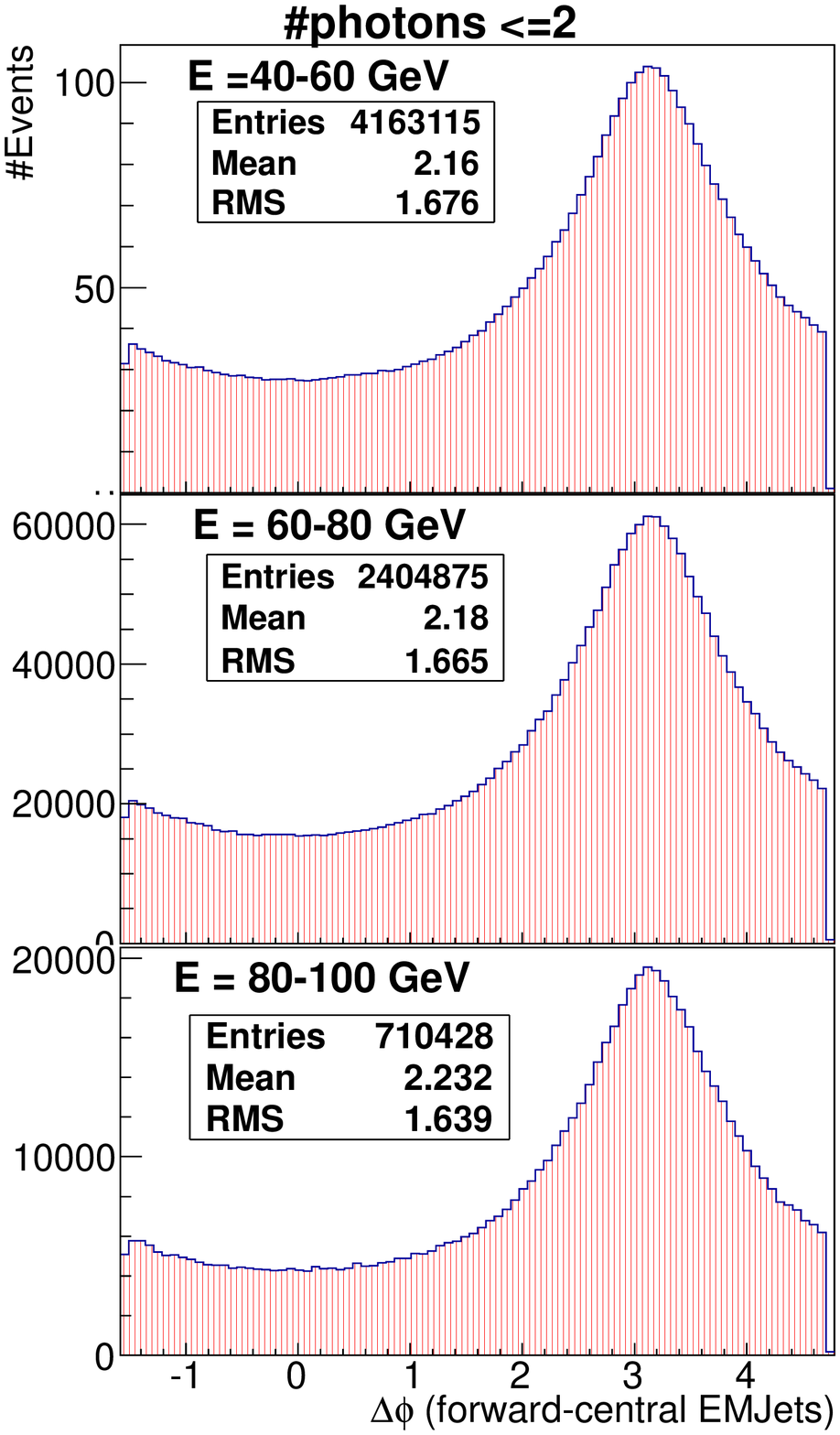}
    \label{fig3}
  \end{subfigure}%
  \begin{subfigure}{0.225\textwidth}
    \centering
    \includegraphics[width=0.98\linewidth]{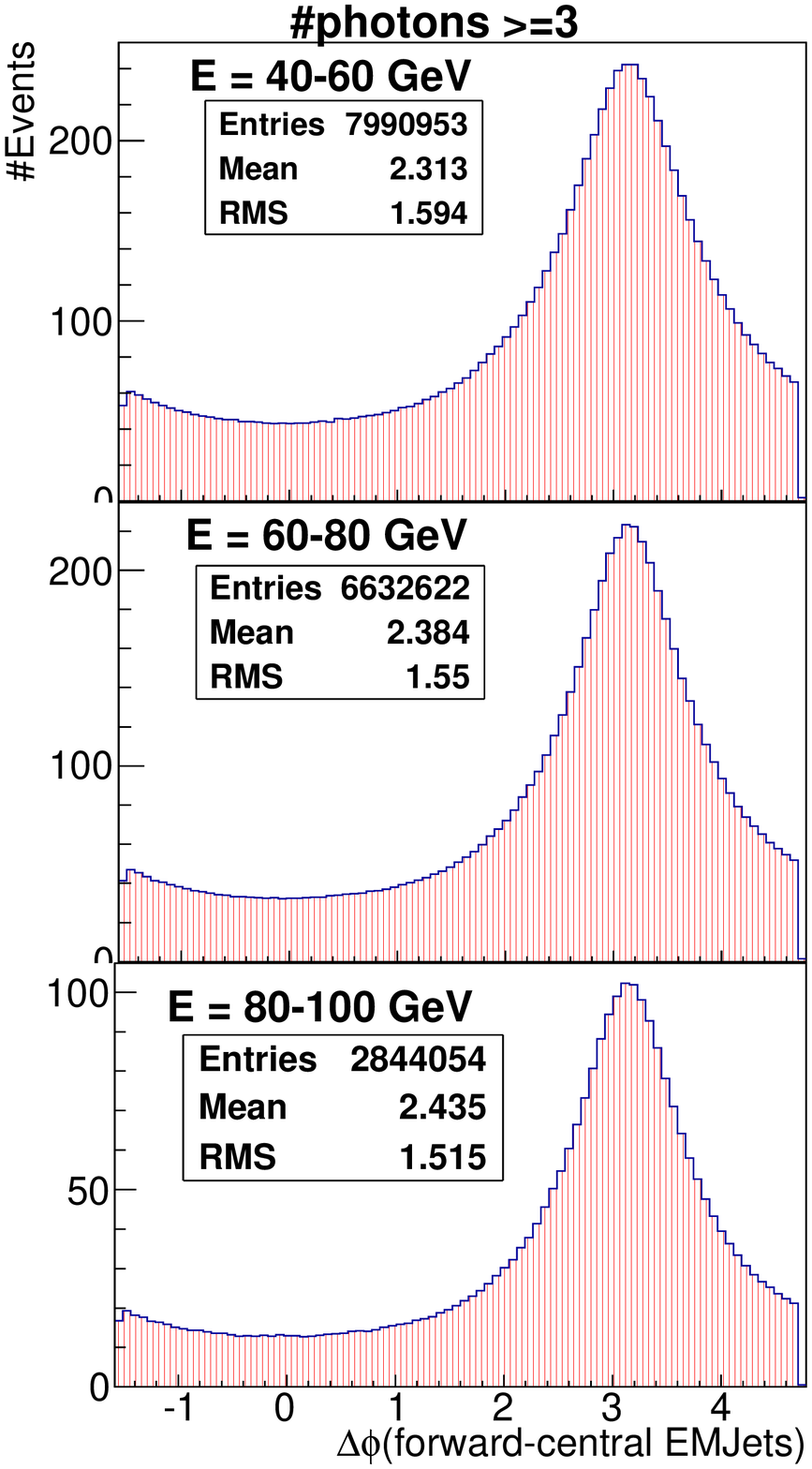}
    \label{fig4}
  \end{subfigure}
  \caption{$\Delta\phi$ correlations between the forward and central EM-Jets for $N_{photons} <=2$ (left)  and for $N_{photons} >2$ (right).}
  \label{F-two}
\end{wrapfigure}
This analysis uses STAR data for {\it p + p} collisions at $\sqrt{s}$ = 500 GeV recorded in 2011 (Run-11)
with 22 pb$^{-1}$ of integrated luminosity and average blue beam (facing the FMS) transverse polarization of 52 $\pm$ 2 $\%$. The trigger is provided 
by the FMS itself (based on the energy deposited in either a cluster of towers or a jet patch), with a threshold to have significant sensitivity to jets and
$\pi^0$'s at high $x_F$ ($=2\times p_{z}/\sqrt{s}$). The trigger from fast detectors limited our analysis
 not to include tracks from central slow detectors like Time Projection Chambers (TPC) .  Neighboring towers in the FMS are grouped into clusters, and then the clusters are fit with characteristic 
electromagnetic shower shape distributions to reconstruct either one or two photons per cluster. The reconstructed photons 
come in large fraction from decays of $\pi^0$ and $\eta$ and with
a small fraction from electrons and direct-$\gamma$. A small fraction of the FMS 
photons also arise from mis-identified hadronic showers in the FMS.

The anti-$k_{T}$ jet clustering algorithm is used with $R$ = 0.7 to find jets separately 
in both the forward (FMS photons) and central (EMC+BEMC towers) rapidity regions, with 
$p_{T}^{EM-jets} > 2.0$ GeV/${\it c}$ and pseudo-rapidity 
$2.8 < \eta^{EM-Jet} < 4.0$ and $-1.0 < \eta^{EM-Jet} < 2.0$, respectively. Only forward EM-jets with 
$40 < E^{EM-Jet} < 100$ GeV are considered, which corresponds the Feynman-$x$ range 
$0.16 < x_{F} < 0.4$.
Only one EM-jet each from the forward and central regions is chosen for an event, the
highest-energy EM-Jet in the forward region and highest-$p_T$ EM-Jet from mid-rapidity.

\begin{wrapfigure}{r}{0.5\textwidth}
  \centering
  \includegraphics[width=.45\textwidth]{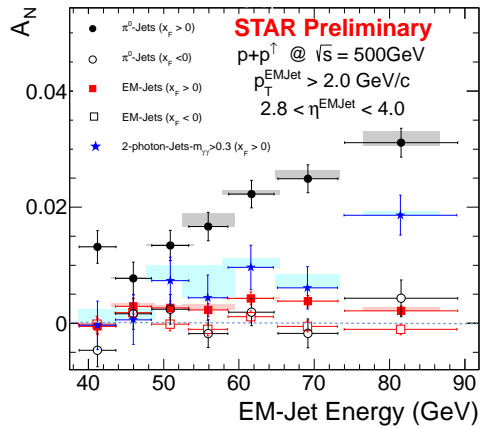}  
  \caption{$A_N$ for EM-jets vs.\@ energy for isolated $\pi^0$, 2-photon EM-jets that fall outside $\pi^0$-mass definition, and for EM-jets containing $N_{photons}$>2.}
  \label{fig3}
\end{wrapfigure} 
The trigger is preferentially sensitive to the neutral energy in the 
forward jets, so the reconstructed forward jets are biased toward those with large 
neutral energy fraction. The number of photons for the forward jets 
lies in the range 1-15 with a most probable value 2. A large fraction of the 1-photon jets arise when the second $\pi^0$-decay 
photon lies outside FMS acceptance or the photon reconstruction mis-identifies a 2-photon cluster as containing only a single $\gamma$.  
Figure \ref{fig2} shows the events with more than two 
photons demonstrate jet-like energy flow, where $\Delta{R_{jet,photons}}$ is the distance in $\eta-\phi$ space between the jet axis and the corresponding photons.
An interesting case for our analysis is the 2-photon case. The invariant mass distribution for 2-photon EM-jets in
Fig.~\ref{fig1} shows a clear $\pi^0$ mass peak.  When the two photons come from a $\pi^0$ decay, they are close together in an otherwise empty jet cone. We define 2-photon EM-Jets having $m_{\gamma\gamma}$ < 0.3 and energy sharing
$z_{\gamma\gamma} = \frac{|E_1-E_2|}{E_1+E_2}$ < 0.8 as isolated-$\pi^0$, or simply ``$\pi^0$-Jets''.  
STAR has previously observed that $A_N$ is larger for isolated-$\pi^0$ than for $\pi^0$ with additional nearby photons~\cite{c8}.
The $\Delta\phi$ correlations between the forward and central EM-Jets in
Fig.~\ref{F-two} show a clear away-side peak which grows stronger with 
increasing forward EM-Jet energies and also with increasing number of photons
in the forward EM-Jets.

The $\phi$ distribution of events for each RHIC fill is fitted with the function
\begin{equation}
  \frac{N{\uparrow}(\phi) - N{\downarrow}(\phi)}{N{\uparrow}(\phi) + N{\downarrow}(\phi)} = p_0 + P\times{A_{N}\cos(\phi)}.
  \label{eq1}
\end{equation}
The values from the L.H.S of 
Eq.~\ref{eq1} are extracted for each of twelve equally divided 
$\phi$-bins.  $P$ is the measured beam polarization for the fill. $p_0$ represents the relative luminosity between spin-up and spin-down collisions for the fill.  The $A_N$ fit results are then averaged over the RHIC fills to obtain the values reported here.

\section{Results}
Figure~\ref{fig3} shows $\pi^0$-Jets have large transverse single-spin 
asymmetries that increase with  energy.  The $A_N$ for $\pi^0$-Jets in Fig.~\ref{fig3} is comparable to that previously reported by STAR for $\pi^0$ that are isolated within cones of 30($\Delta\eta \approx 0.4$) or 70($\Delta\eta \approx 0.7$) mR \cite{c8}.
The asymmetries for EM-jets with more than 2 photons are 
much smaller over the entire energy  range. 
The 2-photon events that do not come 
from $\pi^0$ are expected to arise from $\eta$ decays and 
continuum background.  They have non-zero asymmetries, but much smaller 
than for $\pi^0$-Jets. The systematic errors in Fig.~\ref{fig3} are calculated by 
varying the event cuts and $\eta$ acceptance. 

\begin{figure}
\centering
  \includegraphics[width=0.9\textwidth]{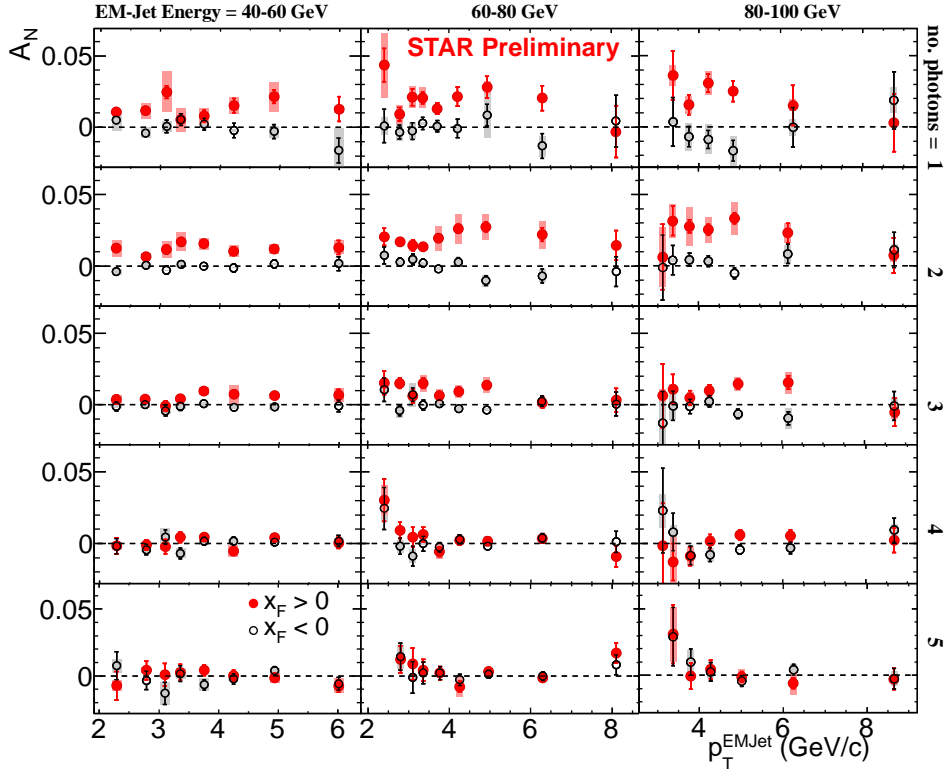}
  \caption{$A_N$ vs.\@ $p_T$ for EM-jets, for $x_F >$ 0 (closed circles) and $x_F <$ 0 (open circles). The three columns correspond to the energy bins 40-60, 60-80, 80-100 GeV.  The five rows indicate different number of photons in the EM-jets from 1 (top) to 5 (bottom).}
  \label{fig4}
\end{figure}
Figure \ref{fig4} shows the $p_T^{EM-Jet}$ dependence of $A_N$ for
$x_F$ > 0 and $x_{F}$ < 0. The results are shown for three energy bins, and
separated according to the number of photons in the EM-jet. $A_N$ for
1-photon events, which include a large $\pi^0$ contribution in this analysis, is
similar to $A_N$ for 2-photon events. Three-photon, jet-like events have a clear non-zero
asymmetry, but substantially smaller than for isolated $\pi^0$'s.  $A_N$ is quite small for EM-jets with 4 or 5 photons. $A_N$ for EM-jets with 6 or more photons (not shown) is very similar to that for 5 photons. It is clear 
that the asymmetries decrease as the complexity, {\it i.e.}, the number of photons
in the event (or the ``jetiness''), increases. The systematic errors shown 
here are calculated from the probability to misidentify the event category, which has been estimated using PYTHIA+GEANT simulations.   

%
\begin{SCfigure}
  \centering
  \includegraphics[width=0.55\textwidth]{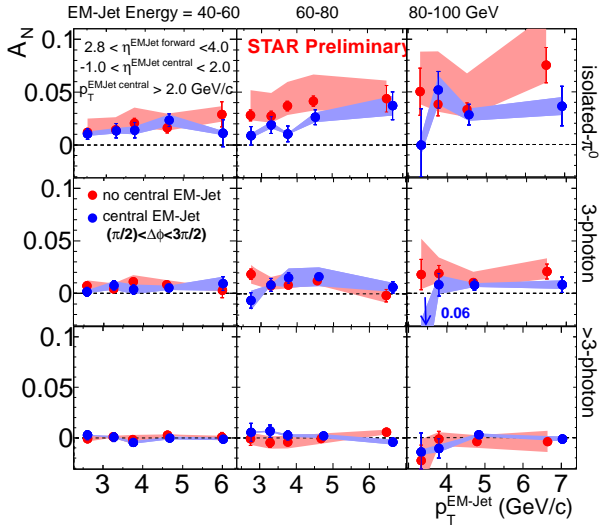}
  \caption{$A_{N}$ vs.\@ $p_T^{EM-Jet}$ for EM-jets when there is a coincident away-side EM-jet at mid-rapidity (blue), and when no coincident away-side EM-jet is seen at mid-rapidity (red). The three columns show energy bins 40-60, 60-80 and 80-100 GeV. The three rows show three different event classes: isolated $\pi^0$, 3-photon EM-jets, and EM-jets with $N_{photons}>3$.}
  \label{fig5}
\end{SCfigure}

Figure \ref{fig5} shows $A_N$ for EM-jets separated according to whether or not a coincident away-side EM-jet is seen at mid-rapidity, where the away side is defined as $\pi/2 < \Delta\phi < 3/2\pi$.  The results are shown for three different event classes and three EM-jet energy bins.  For the isolated-$\pi^0$ case, the asymmetries are smaller
when there is a correlated away-side jet. The systematic errors for 
non-central EM-jets are estimated from the probability to miss a central EM-jet because of finite detector efficiency.
\section{Conclusion}
For the first time, TSSAs are studied for EM-jets in the forward region in $p+p$ collisions at $\sqrt{s}=500$ GeV. We reconfirm that isolated $\pi^0$'s
have large $A_N$. There is a clear non-zero $A_N$ for 3-photon jet-like events, but it is significantly smaller than that for isolated $\pi^0$'s.  $A_N$ then decreases further as the jettiness of the event increases.  Furthermore, we find that $A_N$ for isolated $\pi^0$'s is smaller when they have a coincident away-side EM-jet at mid-rapidity than when no coincident EM-jet is seen.  These results raise serious questions regarding how much
of the large forward $\pi^0$ $A_N$ arises from hard 2$\rightarrow$2 parton 
scatterings.


\begin{thebibliography}{1}
\bibitem{c1} D.L.~Adams {\it et al.} [E581 and E704 Collab.], Phys. Lett. {\bf B261}, 201 (1991); D.L.~Adams {\it et al.} [FNAL-E704 Collab.], Phys. Lett. {\bf B264}, 462 (1991); K.~Krueger {\it et al.}, Phys. Lett. {\bf B459}, 412 (1999). 
\bibitem{c1a} J.~Adams {\it et al.}, Phys. Rev. Lett. {\bf 92}, 171801 (2004) 
\bibitem{c2} D.W.~Sivers, Phys. Rev. {\bf D41}, 83 (1990); Phys. Rev. {\bf D43}, 261 (1991), M.~Anselmino, E.~Murgia, Phys. Lett. {\bf B442}, 470 (1998); U.~D'Alesio, E.~Murgia, Phys. Rev. {\bf D70}, 074009 (2004); M.~Anselmino {\it et al.},Phys. Rev. {\bf D74},094011(2006). 
\bibitem{c4} E.V.~Efremov, O.V.~Teryaev, Sov. J. Nucl. Phys. {\bf 36}, 140 (1982); Phys. Lett. {\bf B150}, 383 (1985), J.W.~Qiu, G.~Sterman, Phys. Rev. Lett. {\bf 67}, 2264 (1991); Nucl. Phys. {\bf B378}, 52 (1992); Phys. Rev. {\bf D59}, 014004 (1999), Y.~Kanazawa, Y.~Koike, Phys. Lett. {\bf B478}, 121 (2000); Phys. Rev. {\bf D64},034019 (2001). 
\bibitem{c6a}Z.~Kang, F.~Yuan, J.~Zhou, Phys. Lett. {\bf b691} 243(2010). 
\bibitem{c7}K.~Ackermann {\it et al.} (STAR Collaboration), Nucl. Inst. \& Meth. {\bf A499}, 624 (2003).
\bibitem{c8} S.~Heppelmann (STAR Collaboration), in {\it Proceedings of the XXI International Workshop on Deep-Inelastic Scattering and Related Subjects (DIS 2013)} (2013).
\end{thebibliography}
\end{document}